# Solar neutrinos and the primordial Sun

Professor Wick Haxton from the University of California Berkeley discusses neutrino experiments past and present and explains how neutrinos may also come to answer the question of whether exo-planet formation leaves chemical imprints on host stars

IN the summer of 1965, the Homestake Mining Company completed the excavation of a 30x60x32 ft cavity on the 4,850 ft level of the Homestake gold mine in Lead, South Dakota, the future site of Ray Davis's chlorine solar neutrino detector. A year later, a massive stainless steel vessel was installed and filled with 100,000 gallons of cleaning fluid, brought to the mine in ten railway tankers. The mile of rock overhead shielded the detector from cosmic rays but left unaffected the electron neutrinos coming from the core of the Sun – by-products of the fusion reactions by which the Sun transforms hydrogen into helium, generating energy.

While in excess of $10^{21}$ solar neutrinos passed through the Davis detector daily, typically only one of these interacted in the tank, converting an atom of chlorine into argon. Because argon is a nonreactive noble gas, Davis was able to design techniques to purge the tank of argon periodically, counting the few atoms of the relevant isotope, $^{37}$Ar, as it decayed back to chlorine.

Davis's team announced their first results in 1968, an upper limit on the solar neutrino flux about one-third that predicted by theory. The detector continued to operate for three decades, confirming what became known as the 'solar neutrino problem'.[1]

## New experiments

Davis's goal in measuring solar neutrinos was to test our understanding of the Sun and other hydrogen-burning stars, which dominate the night sky. The theory of main-sequence stellar evolution, when applied to our best-known star, the Sun, is called the Standard Solar Model (SSM). While the SSM contains about 20 parameters, all of these can be determined from measurements. Constraints included the known mass and radius of the Sun; laboratory measurements of the needed nuclear fusion cross sections and of the atomic processes governing the Sun's radiative opacity; determinations of the depth of the Sun's outer convective zone and of the sound speed in the Sun's interior through helioseismology (the observation and analysis of the fluctuations of the Sun's surface); and measurements of the abundances of various 'metals' in the Sun – elements heavier than helium – which have an important influence on the Sun's radiative opacity.

Once the 20 parameters are fixed, the comparison between predictions and solar neutrino measurements becomes a test of fundamental physics, including weak interaction theory and the physical principles underlying the SSM.

More than two decades passed before new experiments could be performed. But by the early 1990s, results from the radiochemical GALLEX/SAGE and water Cherenkov Kamiokande detectors not only confirmed the deficit that Davis had found, but also revealed a puzzling pattern of solar neutrino fluxes that could not be reproduced by any plausible modification of the SSM. This provided the impetus for a generation of highly sophisticated new experiments, the heavy-water Cerenkov detector SNO (Sudbury Neutrino Observatory), the massive water detector Super-Kamiokande, and the liquid scintillator experiment Borexino.[2]

## Neutrino mass and oscillations

Neutrinos come in three flavours – electron, muon, and tauon – labelled according to the charged leptons with which they are co-produced. Results from SNO had particular impact in resolving the solar neutrino problem, as the detector provided three distinct detection channels for neutrinos, each with a different sensitivity to flavour.

While the Sun generates electron neutrinos, the SNO experimentalists found that their three rates implied that two thirds of the solar neutrinos arriving on Earth were muon or tauon neutrinos – these neutrinos had changed flavour during their eight-minute transit from the Sun's core. This explained Davis's result: his detector was blind to muon and tauon neutrinos.

This result, along with related discoveries made by the Super-Kamiokande collaboration in studies of neutrinos produced by cosmic-ray interactions in the Earth's atmosphere,[3] demonstrate that neutrinos 'oscillate'. This requires that neutrinos have non-zero masses – contrary to the standard model of particle physics – and furthermore, that the states of definite mass are not coincident with the states of definite flavour. In the simplest theory, called the 'seesaw mechanism', the neutrino mass is inversely proportional to the scale of the 'new physics' generating that mass. The solar and atmospheric results showed that neutrinos are very light, reflecting new physics residing at an energy scale far beyond the direct reach of our most powerful accelerators. The deduced scale is close to $M_{GUT}$, the energy at which the fundamental forces – electromagnetic, weak, and strong – may unify, according to theory.

## Problems in the Sun?

The discovery of neutrino mass has had a profound impact on particle physics, stimulating ambitious 'long-baseline' programmes in the US and Japan in which intense accelerator beams of neutrinos are directed toward distant massive detectors.

Yet, the motivation for Davis's experiment was not the discovery of neutrino oscillations, but rather the use of neutrinos as a probe of the solar core. His experiment was primarily sensitive to the highest energy solar neutrinos, those produced in the beta decay of $^8$B. While these neutrinos represent only 0.01% of the total, their flux is acutely sensitive to the solar core temperature, varying as $T_c^{22}$. By detecting these neutrinos, $T_c$ can be determined to an accuracy of ~1%.

About ten years ago, the first crack in the SSM appeared. The model requires one to specify the metallicity of the primordial Sun: the Sun was formed 4.7 Gy ago from a gravitationally contracting gas cloud that was enriched in metals produced in previous generations of stars. As our Sun does not synthesize heavy elements, the primordial metallicity can be determined from observations today of the Sun's surface. Metals generate distinctive absorption lines in the solar spectrum. Alternatively, the metallicity can be determined from the sound speed in the Sun's interior, which can be deduced to sub-1% accuracy from helioseismology.

A fundamental assumption of the SSM is that the primordial Sun was homogeneous. The contracting gas cloud passed through the convective Hayashi phase, where mixing would have destroyed any pre-existing inhomogeneities. Consequently, apart from small corrections due to metal diffusion, the two measurements just described must agree. But they do not, differing by 25% or more.[4]

## Planet formation

This 'solar metallicity problem' has persisted for a decade, despite considerable effort to reconcile the two measurements. An alternative suggestion was made some time ago – that both measurements are correct, reflecting a Sun that is metal-rich in its radiative core but metal-poor in its convective surface region.[5]

The primordial Sun evolved from being fully convective to its modern state, where convection is limited to the outer 2% of the Sun by mass, over a brief period of about 25 My. The Sun's radiative core grew from the inside out, established first in the Sun's centre, then growing in radius. Matter in the radiative core does not mix and is thus isolated from the rest of the Solar System.

This epoch, subsequent to the Hayashi phase, is also the time when the planets formed. On the order of 5% of the original nebular gas then remained, confined to a thin disk that retained most of the solar system's original angular momentum (see Fig. 1). The planets grew by sweeping out the dust and ice concentrated in the disk's midplane, in the process extracting a great quantity of metal from the gas. The concentrations of C and N in the atmospheres of Jupiter and Saturn are four to eight times solar.[6] An estimated 40-90 Earth masses of excess metal are sequestered in the planets.

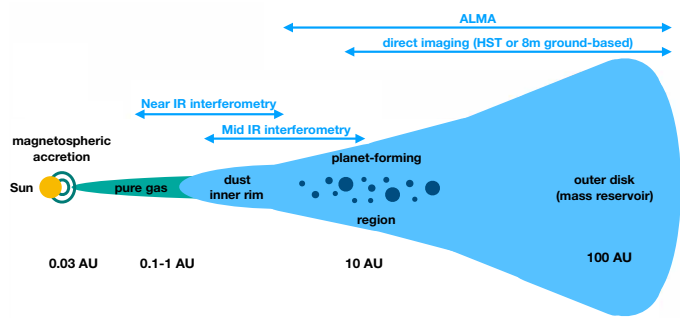

Fig. 1: Artist's depiction of the proto-planetary disk around a young, solar-like star. Adapted with permission from Reference 10

The gas remaining in the disk, hydrogen-rich and scrubbed of its metal, would be expected to accrete onto the Sun. If this accretion occurred when the convective zone resembled its modern form, the resulting dilution could account for the solar metallicity problem.

## Neutrinos and the primordial Sun

This scenario was explored in a self-consistent accreting SSM[5] – a first step in treating the primordial Sun not as an isolate object, but as coupled to the rest of the solar system. As little is known about the detailed composition of the accreting gas, the modelling is subject to many uncertainties.

But this could change. T Tauri stars are recently condensed pre-main-sequence stars resembling the proto-Sun, often with proto-planetary accretion disks. A recent analysis of data from the Gaia space observatory on the inner disks of 26 such stars found very large depletions of carbon – factors of 10-40 – establishing that the gas accreted onto the host star is indeed significantly altered.[7] This result is consistent with recent modelling of accretion disks, which found that volatile molecular species such CO and $CH_4$ are always depleted.[8]

One may be able to use solar neutrinos to directly and precisely[5] measure the metallicity of the Sun's central core. As the first material incorporated into the proto-Sun's evolving radiative zone, the core is the earliest preserved sample of the primordial gas cloud out of which our Solar System formed.

In the CNO cycle, the dominant mechanism for hydrogen burning in stars more massive than our Sun, the nuclear reactions are catalysed by pre-existing metals like C and N. Consequently, the energy production depends linearly on metallicity. While responsible for only 1% of the Sun's energy production, the CNO cycle does operate in the Sun, producing distinctive solar neutrinos. Late last year, the Borexino collaboration, which had already succeeded in measuring all other components of the solar neutrino flux, detected CNO neutrinos.[9] More precise results are expected in the near future, before the experiment ceases running later this year.

These final Borexino results could potentially resolve the solar metallicity problem. If the conclusions drawn from helioseismology are confirmed – that our Sun has a metal-rich core – new solar neutrino experiments to measure the core metallicity as precisely as possible would be needed. There are discussions underway at the new China Jinping Underground Laboratory to mount an experiment very similar to Borexino, but ten times larger and twice (two kilometres!) as deep. If it can be established that the formation of planets affected our Sun's surface metallicity, it would raise interesting questions generally about whether exo-planet formation leaves chemical imprints on host stars.

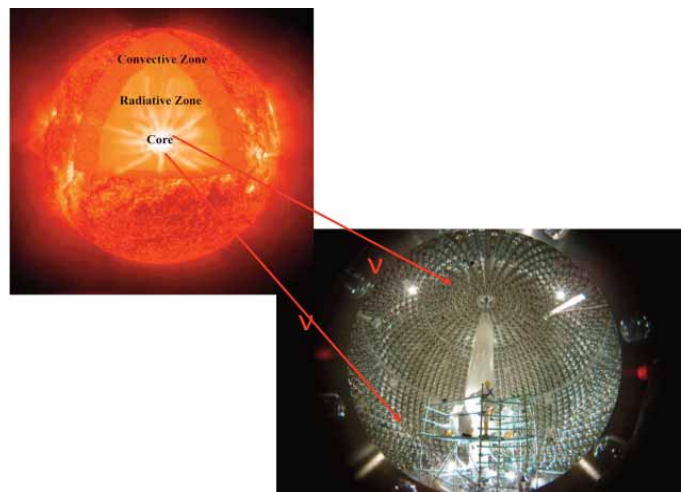

Fig. 2: The Borexino detector, containing ultra-pure liquid scintillator, was designed to measure all of the major components of the solar neutrino flux. It has been operating in the Gran Sasso Laboratory, Italy, since 2007. The collaboration recently detected CNO neutrinos, testing for the first time the mechanism by which massive hydrogen-burning stars produce energy and creating a new opportunity to measure the metallicity of the solar core

**About the author**

Wick Haxton is a Professor of Physics at the University of California, Berkeley, and a Faculty Senior Scientist at Lawrence Berkeley Laboratory. He directs the National Science Foundation Physics Frontier Center N3AS, which focuses on the multi-messenger astrophysics of the Big Bang, supernovae, and merging neutron stars and black holes. This article is based on a public talk he delivered at the 16th Marcel Grossman meeting in July 2021.

WH is supported by the US Department of Energy (grants DE-SC0004658 and DEAC02-05CH11231) and the National Science Foundation (grants 2020275 and 1630782).

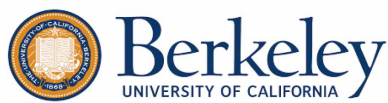

Professor Wick Haxton
Department of Physics
University of California Berkeley
+1 (510) 889 9664

haxton@berkeley.edu
Tweet @BerkeleyPhysics
https://physics.berkeley.edu/
www.facebook.com/UCBerkeleyPhysics